\documentclass[11pt,a4paper]{article}
\usepackage[T1]{fontenc}
\usepackage[utf8]{inputenc}
\usepackage{amsmath,amssymb,graphicx,booktabs,float}
\usepackage{hyperref}
\usepackage[numbers,sort&compress]{natbib}
\usepackage[margin=2.5cm]{geometry}
\usepackage{listings,xcolor}
\newcommand{\pbhalo}{\texttt{pbhalo}}
\newcommand{\Msun}{M_\odot}

\title{\pbhalo: computing primordial black hole and minihalo populations from a single primordial perturbation spectrum}
\author{Bogdan Rudenko\\[2pt]{\small Sternberg Astronomical Institute, Lomonosov Moscow State University}\\{\small\texttt{rudenko.sci@gmail.com}}}
\date{\small To be submitted}

\begin{document}
\maketitle

\begin{abstract}
An enhancement of the primordial curvature power spectrum on small scales can give rise to two populations of compact objects: primordial black holes (PBHs), which form during radiation domination, and dark matter minihalos, including ultracompact minihalos (UCMHs), which form from the same modes during matter domination. We present \pbhalo, an open-source Python package that computes the PBH mass function, the halo mass function and the UCMH abundance from a single spectrum in a given cosmology and applies observational constraints on the PBH abundance and on the spectrum jointly. The package components are validated against \texttt{colossus} (a code for halo mass functions and cosmological quantities), \texttt{pbhstat} (a code for the PBH mass function from a given spectrum) and published results; joint constraints are illustrated for lognormal peaks. The code is distributed under the MIT license.
\end{abstract}

\section{Introduction}
\label{sec:intro}

Primordial black holes, first considered by Zeldovich and Novikov~\citep{ZeldovichNovikov1966}, Hawking~\citep{Hawking1971}, and Carr and Hawking~\citep{CarrHawking1974}, are one of the candidates for a component of dark matter \citep[see the reviews][]{Carr2021,Dokuchaev2026}. The same primordial perturbation modes that produce PBHs can, at lower amplitude, give rise to dark matter minihalos. Moreover, PBHs themselves can act as seeds for dark matter accretion, around which dense halos form, and they can affect the distribution of dark matter in galactic halos \citep{Tkachev2020,Pilipenko2025}. It is therefore natural to treat both populations together, starting from the same primordial spectrum. The aim of this work is a software package that computes both the PBH population and the dark matter minihalo population from a single spectrum in a given cosmology.

The amplitude of the primordial curvature power spectrum $\Delta^2_\zeta(k)$ is measured from the cosmic microwave background and large-scale structure to be $\simeq 2\times10^{-9}$ on scales $10^{-4}\lesssim k/\mathrm{Mpc}^{-1}\lesssim 1$ \citep{Planck2018infl}. Here $\zeta$ is the gauge-invariant curvature perturbation on uniform-density hypersurfaces (conserved outside the horizon), and the dimensionless power spectrum $\Delta^2_\zeta(k)$ is defined by $\langle\zeta_{\mathbf k}\zeta_{\mathbf k'}\rangle=(2\pi)^3\delta_D(\mathbf k+\mathbf k')\,\frac{2\pi^2}{k^3}\Delta^2_\zeta(k)$, so that $\langle\zeta^2(\mathbf x)\rangle=\int\Delta^2_\zeta(k)\,d\ln k$. On smaller scales the amplitude is only bounded from above, and many inflationary scenarios (an inflection point or a plateau in the potential, a phase of ultra-slow roll, multi-field dynamics) predict its enhancement by several orders of magnitude at $k\sim10^{4}$--$10^{15}\,\mathrm{Mpc}^{-1}$ \citep{GarciaBellido2017,MotohashiHu2017,Carr2021,Dokuchaev2026}. Such an enhancement has two observable consequences, which are usually studied separately.

If the enhanced modes have amplitude $\Delta^2_\zeta\gtrsim10^{-2}$, a fraction of horizon-sized regions collapses into PBHs at horizon re-entry during radiation domination; the resulting PBH dark matter fraction $f_{\rm PBH}$ is constrained by evaporation, microlensing, accretion, gravitational waves and dynamical effects \citep[see][]{Carr2021,Kavanagh}. If the amplitude is smaller, $10^{-7}\lesssim\Delta^2_\zeta\lesssim10^{-2}$, no PBHs form, but the same modes grow during matter domination and produce an excess of low-mass dark matter halos collapsing at $z\sim10^{2}$--$10^{3}$: the ultracompact minihalos \citep{Ricotti2009,Bringmann2012,Delos2018}. Their steep density profiles make them bright sources of annihilation radiation and strong lenses, and their non-observation constrains $\Delta^2_\zeta$ at the level $10^{-7}$--$10^{-6}$ over $5\lesssim k/\mathrm{Mpc}^{-1}\lesssim10^{7}$.

Public software covers these two branches separately. \texttt{pbhstat} \citep{pbhstat} computes the PBH mass function from an arbitrary $\Delta^2_\zeta(k)$ with several statistics; \texttt{hmf} \citep{Murray2013}, \texttt{colossus} \citep{Diemer2018} and \texttt{pyccl} \citep{Chisari2019} compute halo mass functions for standard cosmologies but do not accept a primordial spectrum with a small-scale feature as input and do not treat the UCMH regime. To the author's knowledge, there is no public code that computes both populations from one spectrum, describes the transition through matter--radiation equality consistently, and applies the constraints jointly. \pbhalo\ fills this gap.

Treating the two branches together is more than a matter of convenience. First, the amplitudes at which the PBH and UCMH constraints operate differ by four to five orders of magnitude, so the question of which of them sets the allowed region for a given spectral feature cannot be answered without computing both quantities for the same spectrum. Second, both branches are sensitive to the same modelling choices (the smoothing window, the transfer function and the normalisation of the growth factor), and in separate calculations these choices easily become inconsistent. Third, for a spectrum with a narrow peak, constraints formulated for a scale-invariant spectrum cannot be applied directly and must be recomputed for the actual spectral shape; this is required for the extended PBH mass function criterion as well as for the $\mu$ distortion and UCMHs.

\section{Physical model}
\label{sec:physics}

\subsection{Common input}
The input is a function $\Delta^2_\zeta(k)$ with $k$ in comoving Mpc$^{-1}$. Built-in shapes are a power law with running, a lognormal peak
\begin{equation}
\Delta^2_\zeta(k)=\frac{A}{\sqrt{2\pi}\,w}\exp\!\left[-\frac{\ln^2(k/k_p)}{2w^2}\right],\qquad \int\Delta^2_\zeta\,d\ln k = A,
\end{equation}
a broken power law, a table and an arbitrary function; new shapes can be added. All results below are for a Planck 2018 cosmology \citep{Planck2018params} ($h=0.674$, $\Omega_m=0.315$, $\Omega_b=0.0493$), with $z_{\rm eq}=3420$ and $k_{\rm eq}=0.0104\,\mathrm{Mpc}^{-1}$.

\subsection{Radiation-era branch: PBHs}
Following \citet{Gow2021}, the density contrast at horizon entry is $\delta=-\tfrac49(kR)^2\zeta$, so that the variance smoothed on the comoving scale $R$ is
\begin{equation}
\sigma_n^2(R)=\frac{16}{81}\int d\ln k\,k^{2n}(kR)^4W^2(kR)\,T^2(kR)\,\Delta^2_\zeta(k),
\label{eq:sigma}
\end{equation}
where $W$ is a real-space top-hat or Gaussian window, and $T$ is either unity, the linear radiation-era transfer function $T(x)=3(\sin y-y\cos y)/y^3$, $y=x/\sqrt3$, or a sharp cutoff at the first trough of the top hat used by \citet{Gow2021}. The horizon mass of the mode $k$ is $M_H=c^3/2GH(T(k))$ with the Standard Model $g_*(T)$ history of \citet{Husdal2016}; for comparison with the literature, the constant-$g_*$ relation $M_H=M_{\rm eq}(k_{\rm eq}/k)^2/\sqrt2$ is available. PBHs form with mass $M=KM_H(C-C_c)^\gamma$ (critical collapse \citep{Choptuik1993,NiemeyerJedamzik1998}) or $M=\gamma_hM_H$ (threshold model), with Gaussian (Press--Schechter; \citealt{Carr1975}) or BBKS high-peak statistics \citep{BBKS1986}. By default the compaction is non-linear, $C=C_L-\tfrac38C_L^2$, with a Gaussian linear component $C_L<4/3$ \citep[][Appendix C]{Gow2021}. In Press--Schechter statistics, the energy fraction collapsed into PBHs when the scale $R$ enters the horizon is
\begin{equation}
\beta_E(R)=\int_{C_{L,c}}^{4/3}\frac{M(C_L)}{M_H}\,\frac{dC_L}{\sqrt{2\pi}\,\sigma_0}\exp\!\left(-\frac{C_L^2}{2\sigma_0^2}\right),
\label{eq:beta_ps}
\end{equation}
where $C_{L,c}$ is the linear threshold corresponding to $C(C_{L,c})=C_c$, and $\sigma_0$ is the variance~(\ref{eq:sigma}) with $n=0$. In peak statistics, the Gaussian probability density is replaced by the BBKS number density of high peaks per Hubble volume (cf.\ eq.~8 of \citet{Gow2021}):
\begin{equation}
\beta_E(R)=\int_{\nu_c}^{\nu_{\max}}\frac{M(\sigma_0\nu)}{M_H}\,\frac{V(R)}{3^{3/2}(2\pi)^2}\left(\frac{\sigma_1}{\sigma_0}\right)^3\nu^3e^{-\nu^2/2}\,d\nu,\qquad \nu=\frac{C_L}{\sigma_0},
\label{eq:beta_pk}
\end{equation}
where $\sigma_1$ is the moment~(\ref{eq:sigma}) with $n=1$, and $V(R)=\tfrac{4\pi}{3}R^3$ (or $(2\pi)^{3/2}R^3$ for the Gaussian volume); the difference between integrating over $d\nu$ and over $dC_L$ is discussed in Sect.~\ref{sec:validation}. The mass distribution at fixed $R$ is obtained by the change of variable $C_L\to M$ in~(\ref{eq:beta_ps}), after which the contributions of all scales $R$ are summed. The default critical-collapse parameters correspond to the set for which they were calibrated: $K=4$, $C_c=0.55$, $\gamma=0.36$ for a top hat with the large-$k$ cutoff at its first trough and no transfer function, and $K=10$, $C_c=0.25$ for the modified Gaussian window \citep{Gow2021}. The code issues a warning if these values are used with a transfer function or another window. The threshold $C_c$ refers to $C$, so the linear relation $C=C_L$, kept as an option, underestimates the required amplitude. For $C_c=0.55$ the linear threshold is $C_{L,c}=0.776$, and the amplitude giving $f_{\rm PBH}=2\times10^{-3}$ increases by a factor of 1.98--2.00, as found by \citet{Gow2021}. The energy fraction $\beta_E$ collapsed at $a_f$ is converted to the present-day fraction via $\Omega_{\rm PBH}=\beta_E\,\rho(a_f)a_f^3/\rho_{c,0}$ with the $g_*(T)$ and $g_{*s}(T)$ history, and $f_{\rm PBH}=\int f(M)\,d\ln M$.

The threshold $\delta_c$ can be determined from the shape of the spectrum: the code computes the Gaussian shape parameter $\alpha_G$ of the compaction function (eqs.~34 and 37 of \citet{Musco2021}) and their fit $\delta_c(\alpha)$.

\subsection{Matter-era branch: minihalos}
The linear matter power spectrum is
\begin{equation}
P_m(k,z)=\frac{8\pi^2}{25}\frac{(c/H_0)^4}{\Omega_m^2}\,T^2(k)\,k\,\Delta^2_\zeta(k)\,D^2(z),
\label{eq:pm}
\end{equation}
with the Eisenstein--Hu transfer function \citep{EisensteinHu1998} without baryon acoustic oscillations but with baryon suppression (or a CAMB/CLASS table \citep{CAMB,CLASS}), and the growth factor normalised so that $D\to a$ during matter domination. The growth equation is solved including the radiation background from deep in the radiation era, starting from the M\'esz\'aros solution \citep{Meszaros1974} $D=1+\tfrac32a/a_{\rm eq}$; with this normalisation $T(k)$ accounts for the scale dependence of the radiation-era suppression exactly once, and $D(a)$ describes the residual time dependence near $a_{\rm eq}$ that the growth factor of pure $\Lambda$CDM misses (20\% at $z=10^3$, 1\% at $z=50$). The halo mass function is written in the standard form
\begin{equation}
\frac{dn}{d\ln M}=\frac{\rho_m}{M}\,f(\nu)\left|\frac{d\ln\sigma}{d\ln M}\right|,\qquad \nu=\frac{\delta_c(z)}{\sigma(M,z)},
\label{eq:hmf}
\end{equation}
with the Press--Schechter multiplicity function \citep{PressSchechter1974} $f_{\rm PS}=\sqrt{2/\pi}\,\nu\,e^{-\nu^2/2}$ or the Sheth--Tormen one \citep{ShethTormen1999}
\begin{equation}
f_{\rm ST}(\nu)=A_p\sqrt{\frac{2q}{\pi}}\left[1+(q\nu^2)^{-p}\right]\nu\,e^{-q\nu^2/2},
\end{equation}
where $A_p=0.3222$, $p=0.3$; by default $q=1$ \citep{Leo2018}, and the original calibration $q=0.707$ is available as an option. The variance is computed from the spectrum~(\ref{eq:pm}) as $\sigma^2(M,z)=\int d\ln k\,k^3P_m(k,z)W^2(kR)/2\pi^2$. By default $\sigma(M)$ is computed with the smooth-$k$ filter $W=[1+(kR)^{\hat\beta}]^{-1}$, $M=\tfrac{4\pi}{3}\rho_m(\hat cR)^3$, $\hat\beta=4.8$, $\hat c=3.30$, combined with the Sheth--Tormen multiplicity function with $q=1$ \citep{Leo2018}; a sharp-$k$ filter ($c=2.5$; \citealt{Benson2013}) and the real-space top hat are also available. For spectra with a feature this choice matters. The top hat has a $(kR)^{-2}$ envelope at $kR\gg1$, so the power of the peak leaks into $\sigma(M)$ far below the peak mass and produces spurious low-mass halos, the same artefact found for truncated spectra \citep{Schneider2013}. For a peak with $A=10^{-6}$ at $k_p=10^5\,\mathrm{Mpc}^{-1}$ and $z=30$, the top-hat mass function exceeds the result of the $k$-space filters by a factor of 14--38 at $M=10^{-6}$--$10^{-5}\,\Msun$, where the $k$-space filters recover the power-law background. For $\Lambda$CDM, the smooth-$k$ and top-hat mass functions agree to 8\% for $10^{8}\le M/\Msun\le10^{13}$. The logarithmic derivative $d\ln\sigma/d\ln M$ is computed analytically for all filters, and $\delta_c(z)$ is taken from \citet{NakamuraSuto1997}. An optional cutoff $D(k)$ multiplies the transfer function, for example the kinetic-decoupling damping of WIMPs $D=[1-\tfrac23(k/k_{\rm fs})^2]\exp[-(k/k_{\rm fs})^2-(k/k_d)^2]$ \citep{Green2005}. For WIMP dark matter this cutoff falls at $k_{\rm fs}\sim10^{6}\,\mathrm{Mpc}^{-1}$, i.e.\ inside the range of peaks considered here. Two limitations remain. First, the product $T_{\rm EH}(k)D(z)$ is exact only for $z\ll z_{\rm eq}$: at $z\gtrsim300$ the decaying mode of sub-horizon perturbations and the baryons, which do not cluster on small scales and slow the CDM growth to $D\propto a^{0.9}$ until well after recombination, are not described. Second, for $\sigma\gg1$ the mass function is evaluated at $\nu\ll1$, where neither multiplicity function has been calibrated.

\subsection{Ultracompact minihalos}
The UCMH branch implements the approach of \citet{Bringmann2012}: the CDM density variance at horizon entry with the radiation-era CDM transfer function (their eq.~66); the scale-dependent minimum contrast for collapse by $z_c$, obtained from Weinberg's fit \citep{Weinberg2008} to the matter-era transfer function (their eq.~45) and evaluated at the exact cosmic time of the matter--radiation background at $z_c$ (with the Einstein--de~Sitter time the threshold is lower by a factor of 1.7 at $z_c=1200$; the exact time reproduces the values quoted by \citet{Bringmann2025}); the collapse fraction $\beta(k)$ without the Press--Schechter factor of two; the initial mass $M_i=\tfrac{4\pi}{3}\rho_\chi R^3$; growth $M\propto(1+z)^{-1}$ truncated at $z=10$; and the fraction $f_{\rm UCMH}=\beta f_\chi M_0/M_i$. For a given maximum fraction $f_{\max}(M_0)$, this relation is inverted into an upper limit on $\Delta^2_\zeta(k)$ for a locally scale-invariant spectrum ($\sigma^2=0.907\,\Delta^2_\zeta$). For a spectral feature, the code instead computes $f_{\rm UCMH}(k)$ for the actual spectrum and compares its maximum with $f_{\max}$; applying the scale-invariant limit to the peak value of a lognormal peak of width 0.5 would overstate the bound on $A$ by a factor of 1.5--1.8. The annihilation flux is not computed in this paper: $f_{\max}$ is an input parameter. We note that the shallower $r^{-3/2}$ profiles found in simulations \citep{Delos2018} substantially weaken the gamma-ray bound without changing the abundance calculation.

\subsection{Constraints on the PBH fraction of cold dark matter}
The package contains 34 individual bounds on $f_{\rm PBH}(M)$ from PBHbounds \citep{Kavanagh} (evaporation, microlensing, accretion, gravitational waves, dynamics; as listed in the PBHbounds class plots at a fixed commit), as well as the class envelopes for plotting. Extended mass functions are tested with the criterion of \citet{Carr2017}
\begin{equation}
\int d\ln M\,\frac{f(M)}{f_{\max,i}(M)}\le1,
\label{eq:carr}
\end{equation}
where $f_{\max,i}(M)$ is the monochromatic limit from the $i$-th bound. The criterion~(\ref{eq:carr}) is applied to each bound separately: for an envelope $f_{\rm env}=\min_if_i$ the integral exceeds that for any individual bound, so applying the criterion to envelopes over-excludes. The code also reports the fraction of $f_{\rm PBH}$ that lies inside the mass range of at least one bound, since the part of the mass function outside all tables is untested.

The FIRAS limit \citep{Fixsen1996} is applied as the integral constraint
\begin{equation}
\mu=1.401\int dz\,\mathcal{J}_\mu(z)\,\frac{d(Q/\rho_\gamma)}{dz}<9\times10^{-5},
\label{eq:mu}
\end{equation}
with the acoustic-damping heating rate \citep{HuSilk1993,HuScottSilk1994} in the tight-coupling approximation and the $\mu$ visibility function of \citet{Chluba2016} (their eqs.~1, 7b and 11b). For the Planck 2015 spectrum this gives $\mu=2.01\times10^{-8}$, compared with $2.00\times10^{-8}$ for the damping contribution in their Table~1. A tabulated limit on $\Delta^2_\zeta(k)$ is valid only for the spectral shape for which it was derived; the integral form is independent of the shape of the feature. The Planck \citep{Planck2018infl}, Lyman-$\alpha$ forest \citep{Bird2011} and pulsar timing bounds are compared pointwise; the last one \citep{InomataNakama2019} predates the NANOGrav 15-year signal \citep{NANOGrav2023} and is only indicative.

\section{Software}
\label{sec:software}
\pbhalo\ is a pure Python package (NumPy \citep{NumPy}, SciPy \citep{SciPy}). Its modules correspond to the physical blocks of Sect.~\ref{sec:physics}:
\begin{itemize}\setlength\itemsep{0pt}\setlength\parskip{0pt}\setlength\topsep{2pt}
\item \texttt{spectrum}: primordial spectra $\Delta^2_\zeta(k)$: power law, lognormal peak, broken power law, tabulated and arbitrary spectrum; spectra are added with the \texttt{+} operator.
\item \texttt{cosmology}, \texttt{gstar}, \texttt{constants}: background cosmology, solution of the growth equation through equality, mapping between $k$, temperature and horizon mass with the $g_*(T)$, $g_{*s}(T)$ history \citep{Husdal2016}, physical constants.
\item \texttt{transfer}: transfer functions: Eisenstein--Hu (with and without baryons), tabulated from CAMB/CLASS, WIMP and Gaussian cutoffs, and the linear radiation-era transfer function for the PBH branch.
\item \texttt{pbh}: radiation-era branch: moments $\sigma_0$, $\sigma_1$ with different windows, the relation between linear and non-linear compaction, Press--Schechter and peak statistics, critical collapse and threshold model, the mass function $f(M)$ and the total fraction $f_{\rm PBH}$.
\item \texttt{threshold}: threshold $\delta_c$ from the spectral shape: radius $r_m$, shape parameter $\alpha_G$ and the fit of \citet{Musco2021}.
\item \texttt{halo}: matter-era branch: $P_m(k,z)$, $\sigma(M,z)$ for three filters, Press--Schechter and Sheth--Tormen multiplicity functions, halo mass function.
\item \texttt{ucmh}: UCMHs in the formalism of \citet{Bringmann2012}: CDM variance at horizon entry, collapse threshold, fraction $f_{\rm UCMH}(k)$ and its inversion into a limit on the spectrum.
\item \texttt{constraints}: PBHbounds tables and spectrum constraints, criterion~(\ref{eq:carr}) for each bound, coverage of the mass function by the tables, integral $\mu$ distortion.
\item \texttt{pipeline}: the function \texttt{run}, which calls both branches and the constraints for one spectrum and returns the result as a single object.
\item \texttt{plotting}: figures standard in the field (spectrum with excluded regions, $f_{\rm PBH}(M)$ against constraints, mass functions, $\sigma(M)$, UCMH diagnostics, joint-constraint plane) in a uniform style with a colour-blind-safe palette.
\end{itemize}
Units are Mpc, $\Msun$ and comoving Mpc$^{-3}$ without factors of $h$, where $h=H_0/(100\,\mathrm{km\,s^{-1}\,Mpc^{-1}})$ is the dimensionless Hubble parameter.

The central object of the package is the spectrum: any object with a method returning $\Delta^2_\zeta(k)$. Spectra can be added, so the background and the feature are specified independently. The function \texttt{run} computes both branches and applies all constraints; the branches are also available separately through the classes \texttt{PBHMassFunction}, \texttt{HaloMassFunction} and \texttt{UCMH}. A typical session is shown in Listing~\ref{lst:run}.
\begin{lstlisting}[float=!htb,caption={A typical \pbhalo\ session.},label=lst:run]
import numpy as np
from pbhalo import Cosmology, PowerLaw, LognormalPeak, run

spectrum = PowerLaw(A_s=2.1e-9, n_s=0.965) \
         + LognormalPeak(A=9e-3, k_peak=1e6, width=0.5)
res = run(spectrum, Cosmology(), z_halo=(0, 10, 30, 100),
          M_halo=np.geomspace(1e-14, 1e15, 300))
print(res.summary())
res.f_pbh                   # total PBH dark-matter fraction
res.pbh["M"], res.pbh["f"]  # f(M) = d f_PBH / d ln M
res.halo[30]["dn_dlnM"]     # minihalo mass function at z = 30
res.ucmh["f"]               # UCMH fraction per seed scale k
res.excluded                # True if any constraint is violated
\end{lstlisting}

\section{Validation}
\label{sec:validation}
The checks included in the test suite are summarised in Table~\ref{tab:val}.

\begin{table}[!htb]
\centering\small
\begin{tabular}{lll}
\toprule
Quantity & Reference & Agreement\\
\midrule
EH98 transfer function & \texttt{colossus} & $10^{-4}$\\
$\Lambda$CDM growth, $z\le100$ & \texttt{colossus} & $<2\%$\\
Growth rate at $a/a_{\rm eq}=0.1,1,10$ & M\'esz\'aros solution & $<0.5\%$\\
$\sigma_8$ (Planck 2018, EH98) & CAMB 0.811 & 0.836\\
Press--Schechter normalisation (all filters) & analytic & $<1\%$\\
$dn/d\ln M$ in $\Lambda$CDM, smooth-$k$ vs top hat & \citet{Leo2018} & $<8\%$\\
$\mu$ distortion, $\Lambda$CDM & \citet{Chluba2016}: $2.00\times10^{-8}$ & $2.01\times10^{-8}$\\
Horizon mass at $T=1$ MeV & \citet{Carr2021} & $1.50\times10^5\,\Msun$\\
PBH $f(M)$, PS, critical collapse & \texttt{pbhstat} & $10^{-6}$\\
$A$ for $f_{\rm PBH}=2\times10^{-3}$, 8 cases & \citet{Gow2021}, Table~I & $1.7$--$2.0\%$\\
$A_{\rm non\text{-}lin}/A_{\rm lin}$, 4 cases & \citet{Gow2021}, App.~C: 1.98 & 1.977--2.005\\
Threshold for a $\delta$ function: $k r_m$, $\delta_c$ & \citet{Musco2021}: 2.74, 0.51 & 2.743, 0.515\\
Flat spectrum: $k_{\max}r_m$, $\alpha_G$ & \citet{Musco2021}: 4.49, 0.9 & 4.494, 0.901\\
UCMH variance, scale invariance & \citet{Bringmann2012}: $0.907P$ & $<2\%$\\
UCMH masses $M_i$, $M_0$ & \citet{Bringmann2012}, eqs.~1, 3 & $<2\%$, $<15\%$\\
\bottomrule
\end{tabular}
\caption{Checks included in the test suite.}
\label{tab:val}
\end{table}

The 3\% excess of $\sigma_8$ over CAMB corresponds to the known accuracy of the Eisenstein--Hu fit (and to neglecting massive neutrinos); a tabulated transfer function removes it. The amplitudes of Gow et al.\ are reproduced in their conventions, which differ from the \pbhalo\ defaults in two places. First, their eq.~(8) integrates the peak number density $n(\nu)$ over $dC$ rather than over $d\nu=dC/\sigma_0$; \pbhalo\ includes the Jacobian by default, which increases $\beta$ by a factor of $1/\sigma_0\simeq13$--$30$ at the relevant amplitudes and lowers the amplitude required for a given $f_{\rm PBH}$ by 10--13\%. Second, their conversion $f=\beta(M_{\rm eq}/M_H)^{1/2}/\Omega_{\rm cdm}$ exceeds the dilution computed from the thermal history by a factor of $1/(\sqrt2\,\Omega_m)\simeq2.2$ for constant $g_*$ and up to 3.6 for the Standard Model $g_*(T)$, i.e.\ by 4--7\% in amplitude. With both conventions switched to theirs and with the linear compaction $C=C_L$ assumed in their Table~I, the eight cases (two windows, two statistics, two peak widths) agree to 1.7--2.0\%, with a common offset that is probably related to the definition of the peak scale. Since $d\ln f_{\rm PBH}/d\ln A\simeq30$, a 2\% difference in $A$ corresponds to a factor of $\simeq1.8$ in $f_{\rm PBH}$.

\section{Example: a lognormal peak at $k_p=10^6\,\mathrm{Mpc}^{-1}$}
\label{sec:example}

\begin{figure}[!htb]
\centering
\includegraphics[width=0.9\textwidth]{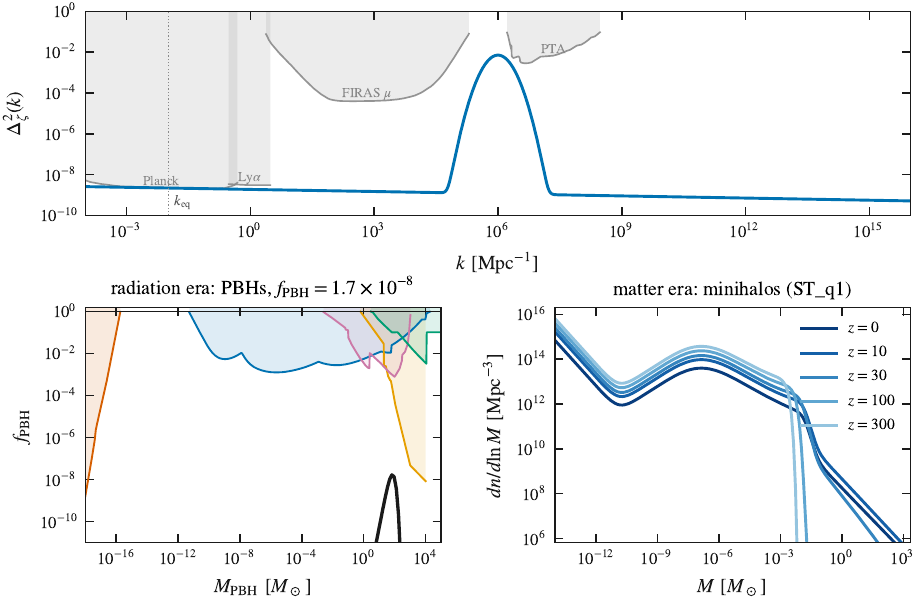}
\caption{The two branches for a power-law spectrum with a lognormal peak $A=9\times10^{-3}$, $k_p=10^6\,\mathrm{Mpc}^{-1}$, $w=0.5$. Top: the spectrum and the constraints on it. Bottom left: the PBH mass function (black line, $f_{\rm PBH}=1.7\times10^{-8}$) against the PBHbounds class constraints. Bottom right: the minihalo mass function (Sheth--Tormen, $q=1$) at $z=0$--$300$.}
\label{fig:two}
\end{figure}

To illustrate both branches, consider a Planck spectrum with a power-law background ($A_s=2.1\times10^{-9}$, $n_s=0.965$) to which a lognormal peak with $A=9\times10^{-3}$, $k_p=10^6\,\mathrm{Mpc}^{-1}$ and $w=0.5$ is added (Fig.~\ref{fig:two}, top). Such a peak does not violate any of the constraints on $\Delta^2_\zeta$ directly: it lies to the right of the FIRAS sensitivity region and to the left of the pulsar timing region. However, it falls deep inside the region excluded by the UCMH bound if dark matter annihilates.

\begin{figure}[!htb]
\centering
\includegraphics[width=0.48\textwidth]{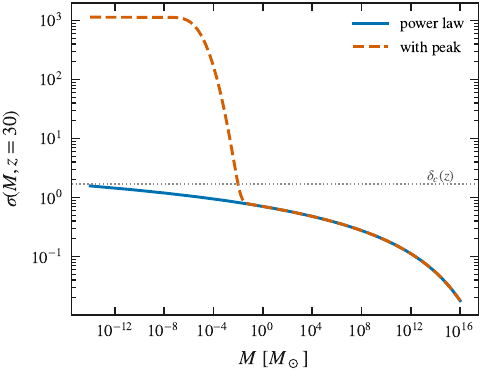}\hfill
\includegraphics[width=0.48\textwidth]{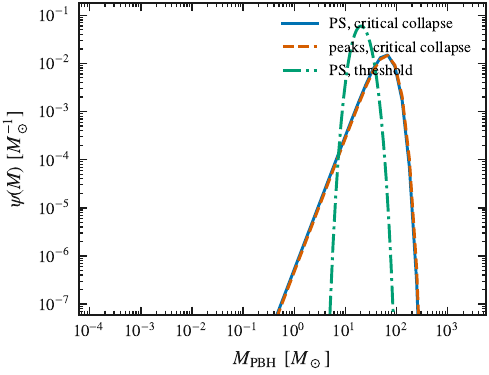}
\caption{Left: the variance $\sigma(M)$ at $z=30$ for the power-law spectrum and the spectrum with a peak (smooth-$k$ filter); the dashed line is $\delta_c(z)$. Right: the normalised PBH mass function $\psi(M)$ for Press--Schechter and peak statistics with critical collapse and for the threshold model.}
\label{fig:sigma_pbh}
\end{figure}

The result of computing both branches is summarised in Fig.~\ref{fig:two} (bottom). In the radiation-era branch the peak produces a small but computable PBH population: the total fraction in Press--Schechter statistics is $f_{\rm PBH}=1.7\times10^{-8}$, and the mass function is concentrated around $10$--$100\,\Msun$, i.e.\ near the horizon mass for $k_p$. In peak statistics the same amplitude gives $f_{\rm PBH}=1.6\times10^{-5}$. This fraction is several orders of magnitude below the PBHbounds constraints in this mass range. Owing to the exponential dependence of $\beta$ on $\sigma_0$, reducing the amplitude by only a factor of 1.8, to $A=5\times10^{-3}$, lowers $f_{\rm PBH}$ to $\sim10^{-19}$, while $f_{\rm PBH}\sim1$ requires an amplitude only about twice as large, $A\approx1.9\times10^{-2}$ (see Sect.~\ref{sec:application}). The shape of the normalised mass function (Fig.~\ref{fig:sigma_pbh}, right) is nearly independent of the choice between Press--Schechter and peak statistics, whereas the threshold model $M=\gamma_hM_H$ gives a narrower distribution shifted to lower masses; critical collapse produces the characteristic power-law tail at low masses.

\begin{figure}[!htb]
\centering
\includegraphics[width=0.45\textwidth]{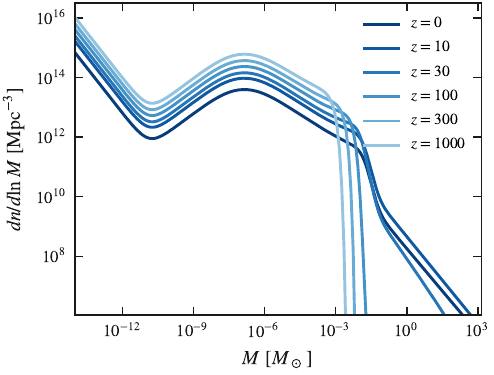}
\caption{The minihalo mass function for the spectrum with a peak at $z=0$, 10, 30, 100, 300 and 1000.}
\label{fig:hmf}
\end{figure}

In the matter-era branch the same peak shows up strongly. The variance $\sigma(M)$ at $z=30$ (Fig.~\ref{fig:sigma_pbh}, left) rises from values of order unity to $\sim10^3$ at masses below $\sim10^{-3}\,\Msun$ and reaches a plateau corresponding to the total power of the peak. Since $\sigma\gg\delta_c$, practically all matter on these scales has collapsed by $z=30$, and the halo mass function (Fig.~\ref{fig:hmf}) has a broad bump around $10^{-7}\,\Msun$, exceeding the background by several orders of magnitude. As $z$ decreases the bump goes down, because minihalos merge into more massive objects, and the sharp high-mass cutoff moves to the right. As noted in Sect.~\ref{sec:physics}, for $\sigma\gg1$ the multiplicity functions operate outside their calibration range, so the shape of the bump is reliable only qualitatively; the smooth-$k$ filter, however, guarantees the absence of the spurious excess of halos below the peak mass that is characteristic of the top-hat window.

\begin{figure}[!htb]
\centering
\includegraphics[width=0.85\textwidth]{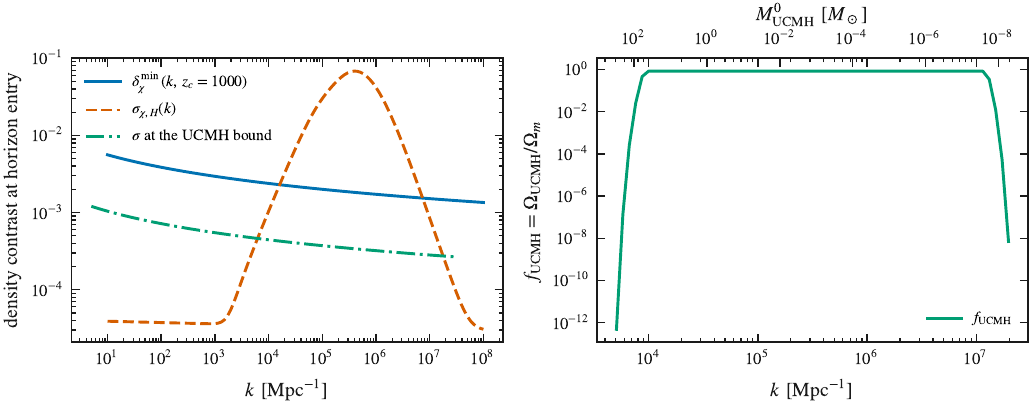}
\caption{UCMH diagnostics for the spectrum with a peak. Left: the minimum CDM density contrast at horizon entry for collapse by $z_c=1000$ (solid), the variance $\sigma_{\chi,H}(k)$ for the spectrum with a peak (dashed), and the variance corresponding to the bound $f_{\rm UCMH}<10^{-5}$ (dash-dotted). Right: the UCMH fraction $f_{\rm UCMH}(k)$; the top axis gives the present-day UCMH mass.}
\label{fig:ucmh}
\end{figure}

The UCMH diagnostics are shown in Fig.~\ref{fig:ucmh}. The CDM density variance at horizon entry $\sigma_{\chi,H}(k)$ exceeds the minimum contrast for collapse by $z_c=1000$ over $2\times10^{4}\lesssim k/\mathrm{Mpc}^{-1}\lesssim7\times10^{6}$, and in this interval the UCMH fraction saturates, $f_{\rm UCMH}\simeq1$. For comparison, the bound $f_{\rm UCMH}<10^{-5}$ requires $\sigma_{\chi,H}$ to lie below the dash-dotted line, i.e.\ about four orders of magnitude below the peak variance in terms of $\Delta^2_\zeta$. Thus a peak that is entirely safe from the point of view of PBHs and spectral distortions is excluded by a huge margin if dark matter annihilates, and it is this asymmetry that determines the structure of the joint constraints in the next section.

\section{Application: joint constraints on a lognormal peak}
\label{sec:application}

\begin{figure}[!htb]
\centering
\includegraphics[width=0.7\textwidth]{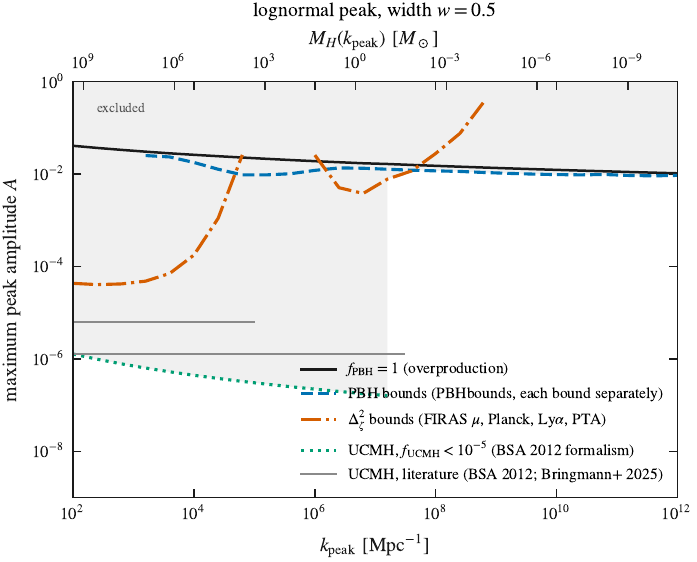}
\caption{Maximum amplitude $A$ of a lognormal peak of width $w=0.5$ as a function of its position $k_p$. Black line: PBH overproduction ($f_{\rm PBH}=1$). Blue dashed: observational PBH constraints, each tested separately with the extended-mass-function criterion; shown where at least 90\% of $f_{\rm PBH}$ lies within the mass range of some bound. Orange dash-dotted: constraints on $\Delta^2_\zeta$ (FIRAS $\mu$ integral, Planck, Lyman-$\alpha$, PTA); gaps mark peak positions not constrained by any of them. Green dotted: the UCMH bound for a maximum UCMH fraction $f_{\rm UCMH}<10^{-5}$ in the Bringmann et al.\ formalism, computed for the actual spectrum; grey lines: approximate UCMH bounds from the literature, the gamma-ray bound of \citet{Bringmann2012} and the annihilation-independent bound of \citet{Bringmann2025}. Top axis: horizon mass for the peak scale.}
\label{fig:joint}
\end{figure}

Figure~\ref{fig:joint} shows the maximum peak amplitude allowed by each class of constraints. Four features are worth noting (for non-linear compaction and calibrated critical-collapse parameters; the linear relation would lower the PBH limits by a factor of $\simeq2$). First, at $1.6\times10^{3}\lesssim k_p/\mathrm{Mpc}^{-1}\lesssim10^{12}$ the PBH constraints limit the amplitude to $A\simeq(0.9$--$2.5)\times10^{-2}$, a factor of 1.1--2.4 below the overproduction line; the exponential sensitivity of $f_{\rm PBH}$ to $A$ makes the observational constraints nearly equivalent to the requirement $f_{\rm PBH}<1$. At smaller $k_p$ the PBH masses fall outside the range of the tabulated constraints. Applying the Carr et al.\ criterion to class envelopes instead of individual bounds would tighten the limit on $A$ by up to a factor of 10 (at $k_p\sim10^3\,\mathrm{Mpc}^{-1}$). Second, the $\mu$ distortion limits the peak to $A\lesssim(4$--$5)\times10^{-5}$ at $k_p\lesssim2\times10^{3}\,\mathrm{Mpc}^{-1}$. Above $10^{4}\,\mathrm{Mpc}^{-1}$ the constraint weakens rapidly, because such modes are damped before the thermalisation redshift, and it disappears at $k_p\gtrsim10^{5}\,\mathrm{Mpc}^{-1}$. Pulsar timing reaches $A\simeq4\times10^{-3}$ at $k_p\simeq6\times10^{6}\,\mathrm{Mpc}^{-1}$. Third, if dark matter annihilates, the UCMH bound gives $A\lesssim(1.6$--$13)\times10^{-7}$ at $10^{2}\le k_p/\mathrm{Mpc}^{-1}\lesssim1.6\times10^{7}$. This is stronger than the next strongest constraint by a factor of 34 at $k_p=10^2\,\mathrm{Mpc}^{-1}$ (where the $\mu$ limit applies) and by up to $5\times10^{4}$ at larger $k_p$. The UCMH bound depends only weakly on the adopted $f_{\max}$: changing $f_{\max}$ from $10^{-7}$ to $10^{-3}$ shifts it by a factor of 1.9. Fourth, these UCMH numbers rely on the Bringmann et al.\ formalism, whose density profiles and accretion history have been revised on the basis of simulations \citep{Delos2018}, and on the adopted $f_{\max}$; they should be understood as the sensitivity of the annihilation signal rather than as a robust limit. With these caveats, a spectral feature that produces an interesting PBH population without violating the UCMH bound requires either non-annihilating dark matter or a peak at $k_p\gtrsim3\times10^{7}\,\mathrm{Mpc}^{-1}$, i.e.\ PBH masses below $\sim10^{-2}\,\Msun$. This complementarity, which is the main motivation for computing both branches from one spectrum, is obtained here with a single function call per peak position.

\section{Discussion and outlook}
\label{sec:discussion}

\pbhalo\ is intended as a tool in which all modelling choices are explicit and consistent between the branches, rather than as the definitive model of each of them. The main sources of systematic uncertainty relevant for applications are as follows.

\emph{PBH formation.} The collapse threshold, the critical-collapse parameters, the choice of window and the relation between linear and non-linear compaction change the amplitude required for a given $f_{\rm PBH}$ by tens of per cent, which, owing to the steep dependence $f_{\rm PBH}(A)$, translates into orders of magnitude in $f_{\rm PBH}$. In \pbhalo\ the defaults are fixed to the sets for which the parameters were calibrated \citep{Gow2021}, and mixing them triggers a warning. Non-Gaussianity, changes of the equation of state at early-Universe phase transitions, and PBH clustering are not included.

\emph{UCMHs.} The Bringmann et al.\ formalism assumes steep profiles $\rho\propto r^{-9/4}$ and mass growth by secondary accretion; simulations \citep{Delos2018} point to shallower profiles and a smaller role of UCMHs as a distinct population. Since $f_{\max}$ is an input parameter, the package allows any signal model to be plugged in, but the interpretation of the UCMH bound remains model-dependent.

The code, tests, documentation and the scripts producing the figures of this paper are available at \url{https://github.com/RudenkoBogdan/pbhalo} under the MIT license; the constraint tables are redistributed from PBHbounds under the same license. Planned extensions include the non-linear mapping $\alpha_G\to\alpha$ of \citet{Musco2021} and computation of the annihilation flux from UCMHs with the \citet{Delos2018} profiles.

\section*{Acknowledgements}
The author thanks K.\,A.~Postnov and N.\,K.~Porayko for useful discussions. This work was supported by a scholarship of the ``Frontiers of Science'' foundation.

\bibliographystyle{unsrtnat}
\bibliography{refs}

\begin{thebibliography}{49}
\providecommand{\natexlab}[1]{#1}
\providecommand{\url}[1]{\texttt{#1}}
\expandafter\ifx\csname urlstyle\endcsname\relax
  \providecommand{\doi}[1]{doi: #1}\else
  \providecommand{\doi}{doi: \begingroup \urlstyle{rm}\Url}\fi

\bibitem[Zel'dovich and Novikov(1966)]{ZeldovichNovikov1966}
Ya.~B. Zel'dovich and I.~D. Novikov.
\newblock The hypothesis of cores retarded during expansion and the hot
  cosmological model.
\newblock \emph{Astron. Zh.}, 43:\penalty0 758, 1966.
\newblock Sov. Astron. 10, 602 (1967).

\bibitem[Hawking(1971)]{Hawking1971}
Stephen Hawking.
\newblock Gravitationally collapsed objects of very low mass.
\newblock \emph{Mon. Not. R. Astron. Soc.}, 152:\penalty0 75, 1971.

\bibitem[Carr and Hawking(1974)]{CarrHawking1974}
Bernard~J. Carr and Stephen~W. Hawking.
\newblock Black holes in the early universe.
\newblock \emph{Mon. Not. R. Astron. Soc.}, 168:\penalty0 399, 1974.

\bibitem[Carr et~al.(2021)Carr, Kohri, Sendouda, and Yokoyama]{Carr2021}
Bernard Carr, Kazunori Kohri, Yuuiti Sendouda, and Jun'ichi Yokoyama.
\newblock Constraints on primordial black holes.
\newblock \emph{Rep. Prog. Phys.}, 84:\penalty0 116902, 2021.

\bibitem[Dokuchaev et~al.(2026)Dokuchaev, Eroshenko, Nikulin, Postnov, Rubin,
  and Stasenko]{Dokuchaev2026}
V.~I. Dokuchaev, Yu.~N. Eroshenko, V.~V. Nikulin, K.~A. Postnov, S.~G. Rubin,
  and V.~D. Stasenko.
\newblock Primordial black holes.
\newblock \emph{Usp. Fiz. Nauk}, 196:\penalty0 1011--1054, 2026.
\newblock \doi{10.3367/UFNr.2026.07.040164}.
\newblock arXiv:2610.10750.

\bibitem[Tkachev et~al.(2020)Tkachev, Pilipenko, and Yepes]{Tkachev2020}
M.~V. Tkachev, S.~V. Pilipenko, and G.~Yepes.
\newblock Dark matter simulations with primordial black holes in the early
  universe.
\newblock \emph{Mon. Not. R. Astron. Soc.}, 499:\penalty0 4854, 2020.

\bibitem[Pilipenko et~al.(2025)Pilipenko, Tkachev, and
  Arakelyan]{Pilipenko2025}
S.~V. Pilipenko, M.~V. Tkachev, and N.~R. Arakelyan.
\newblock On the increase in the concentration of primordial black holes in the
  halos of dwarf galaxies.
\newblock \emph{JETP Lett.}, 121:\penalty0 324, 2025.

\bibitem[{Planck Collaboration} et~al.(2020{\natexlab{a}}){Planck
  Collaboration}, Akrami, et~al.]{Planck2018infl}
{Planck Collaboration}, Y.~Akrami, et~al.
\newblock Planck 2018 results. x. constraints on inflation.
\newblock \emph{Astron. Astrophys.}, 641:\penalty0 A10, 2020{\natexlab{a}}.

\bibitem[Garc{\'i}a-Bellido and Ruiz~Morales(2017)]{GarciaBellido2017}
Juan Garc{\'i}a-Bellido and Ester Ruiz~Morales.
\newblock Primordial black holes from single field models of inflation.
\newblock \emph{Phys. Dark Univ.}, 18:\penalty0 47, 2017.

\bibitem[Motohashi and Hu(2017)]{MotohashiHu2017}
Hayato Motohashi and Wayne Hu.
\newblock Primordial black holes and slow-roll violation.
\newblock \emph{Phys. Rev. D}, 96:\penalty0 063503, 2017.

\bibitem[Kavanagh(2019)]{Kavanagh}
Bradley~J. Kavanagh.
\newblock Pbhbounds.
\newblock \url{https://github.com/bradkav/PBHbounds}, 2019.
\newblock DOI 10.5281/zenodo.3538999.

\bibitem[Ricotti and Gould(2009)]{Ricotti2009}
Massimo Ricotti and Andrew Gould.
\newblock A new probe of dark matter and high-energy universe using
  microlensing.
\newblock \emph{Astrophys. J.}, 707:\penalty0 979, 2009.

\bibitem[Bringmann et~al.(2012)Bringmann, Scott, and Akrami]{Bringmann2012}
Torsten Bringmann, Pat Scott, and Yashar Akrami.
\newblock Improved constraints on the primordial power spectrum at small scales
  from ultracompact minihalos.
\newblock \emph{Phys. Rev. D}, 85:\penalty0 125027, 2012.

\bibitem[Delos et~al.(2018)Delos, Erickcek, Bailey, and Alvarez]{Delos2018}
M.~Sten Delos, Adrienne~L. Erickcek, Avery~P. Bailey, and Marcelo~A. Alvarez.
\newblock Density profiles of ultracompact minihalos: Implications for
  constraining the primordial power spectrum.
\newblock \emph{Phys. Rev. D}, 98:\penalty0 063527, 2018.

\bibitem[Cole and Fumagalli(2025)]{pbhstat}
Philippa~S. Cole and Jacopo Fumagalli.
\newblock pbhstat: a python code for the primordial black hole mass function
  and abundance.
\newblock \url{https://github.com/pipcole/pbhstat}, 2025.

\bibitem[Murray et~al.(2013)Murray, Power, and Robotham]{Murray2013}
Steven~G. Murray, Chris Power, and Aaron S.~G. Robotham.
\newblock Hmfcalc: An online tool for calculating dark matter halo mass
  functions.
\newblock \emph{Astron. Comput.}, 3:\penalty0 23, 2013.

\bibitem[Diemer(2018)]{Diemer2018}
Benedikt Diemer.
\newblock Colossus: A python toolkit for cosmology, large-scale structure, and
  dark matter halos.
\newblock \emph{Astrophys. J. Suppl.}, 239:\penalty0 35, 2018.

\bibitem[Chisari et~al.(2019)]{Chisari2019}
Nora~Elisa Chisari et~al.
\newblock Core cosmology library: Precision cosmological predictions for lsst.
\newblock \emph{Astrophys. J. Suppl.}, 242:\penalty0 2, 2019.

\bibitem[{Planck Collaboration} et~al.(2020{\natexlab{b}}){Planck
  Collaboration}, Aghanim, et~al.]{Planck2018params}
{Planck Collaboration}, N.~Aghanim, et~al.
\newblock Planck 2018 results. vi. cosmological parameters.
\newblock \emph{Astron. Astrophys.}, 641:\penalty0 A6, 2020{\natexlab{b}}.

\bibitem[Gow et~al.(2021)Gow, Byrnes, Cole, and Young]{Gow2021}
Andrew~D. Gow, Christian~T. Byrnes, Philippa~S. Cole, and Sam Young.
\newblock The power spectrum on small scales: robust constraints and comparing
  pbh methodologies.
\newblock \emph{JCAP}, 02:\penalty0 002, 2021.

\bibitem[Husdal(2016)]{Husdal2016}
Lars Husdal.
\newblock On effective degrees of freedom in the early universe.
\newblock \emph{Galaxies}, 4:\penalty0 78, 2016.

\bibitem[Choptuik(1993)]{Choptuik1993}
Matthew~W. Choptuik.
\newblock Universality and scaling in gravitational collapse of a massless
  scalar field.
\newblock \emph{Phys. Rev. Lett.}, 70:\penalty0 9, 1993.

\bibitem[Niemeyer and Jedamzik(1998)]{NiemeyerJedamzik1998}
J.~C. Niemeyer and K.~Jedamzik.
\newblock Near-critical gravitational collapse and the initial mass function of
  primordial black holes.
\newblock \emph{Phys. Rev. Lett.}, 80:\penalty0 5481, 1998.

\bibitem[Carr(1975)]{Carr1975}
Bernard~J. Carr.
\newblock The primordial black hole mass spectrum.
\newblock \emph{Astrophys. J.}, 201:\penalty0 1, 1975.

\bibitem[Bardeen et~al.(1986)Bardeen, Bond, Kaiser, and Szalay]{BBKS1986}
J.~M. Bardeen, J.~R. Bond, N.~Kaiser, and A.~S. Szalay.
\newblock The statistics of peaks of gaussian random fields.
\newblock \emph{Astrophys. J.}, 304:\penalty0 15, 1986.

\bibitem[Musco et~al.(2021)Musco, De~Luca, Franciolini, and Riotto]{Musco2021}
Ilia Musco, Valerio De~Luca, Gabriele Franciolini, and Antonio Riotto.
\newblock The threshold for primordial black hole formation: a simple analytic
  prescription.
\newblock \emph{Phys. Rev. D}, 103:\penalty0 063538, 2021.

\bibitem[Eisenstein and Hu(1998)]{EisensteinHu1998}
Daniel~J. Eisenstein and Wayne Hu.
\newblock Baryonic features in the matter transfer function.
\newblock \emph{Astrophys. J.}, 496:\penalty0 605, 1998.

\bibitem[Lewis et~al.(2000)Lewis, Challinor, and Lasenby]{CAMB}
Antony Lewis, Anthony Challinor, and Anthony Lasenby.
\newblock Efficient computation of cmb anisotropies in closed frw models.
\newblock \emph{Astrophys. J.}, 538:\penalty0 473, 2000.

\bibitem[Blas et~al.(2011)Blas, Lesgourgues, and Tram]{CLASS}
Diego Blas, Julien Lesgourgues, and Thomas Tram.
\newblock The cosmic linear anisotropy solving system (class). part ii:
  Approximation schemes.
\newblock \emph{JCAP}, 07:\penalty0 034, 2011.

\bibitem[M{\'e}sz{\'a}ros(1974)]{Meszaros1974}
P.~M{\'e}sz{\'a}ros.
\newblock The behaviour of point masses in an expanding cosmological
  substratum.
\newblock \emph{Astron. Astrophys.}, 37:\penalty0 225, 1974.

\bibitem[Press and Schechter(1974)]{PressSchechter1974}
William~H. Press and Paul Schechter.
\newblock Formation of galaxies and clusters of galaxies by self-similar
  gravitational condensation.
\newblock \emph{Astrophys. J.}, 187:\penalty0 425, 1974.

\bibitem[Sheth and Tormen(1999)]{ShethTormen1999}
Ravi~K. Sheth and Giuseppe Tormen.
\newblock Large-scale bias and the peak background split.
\newblock \emph{Mon. Not. R. Astron. Soc.}, 308:\penalty0 119, 1999.

\bibitem[Leo et~al.(2018)Leo, Baugh, Li, and Pascoli]{Leo2018}
Matteo Leo, Carlton~M. Baugh, Baojiu Li, and Silvia Pascoli.
\newblock A new smooth-k space filter approach to calculate halo abundances.
\newblock \emph{JCAP}, 04:\penalty0 010, 2018.
\newblock \doi{10.1088/1475-7516/2018/04/010}.

\bibitem[Benson et~al.(2013)]{Benson2013}
Andrew~J. Benson et~al.
\newblock Dark matter halo merger histories beyond cold dark matter -- i.
  methods and application to warm dark matter.
\newblock \emph{Mon. Not. R. Astron. Soc.}, 428:\penalty0 1774, 2013.

\bibitem[Schneider et~al.(2013)Schneider, Smith, and Reed]{Schneider2013}
Aurel Schneider, Robert~E. Smith, and Darren Reed.
\newblock Halo mass function and the free streaming scale.
\newblock \emph{Mon. Not. R. Astron. Soc.}, 433:\penalty0 1573, 2013.
\newblock \doi{10.1093/mnras/stt829}.

\bibitem[Nakamura and Suto(1997)]{NakamuraSuto1997}
Takahiro~T. Nakamura and Yasushi Suto.
\newblock Strong gravitational lensing and velocity function as tools to probe
  cosmological parameters.
\newblock \emph{Prog. Theor. Phys.}, 97:\penalty0 49, 1997.

\bibitem[Green et~al.(2005)Green, Hofmann, and Schwarz]{Green2005}
Anne~M. Green, Stefan Hofmann, and Dominik~J. Schwarz.
\newblock The first wimpy halos.
\newblock \emph{JCAP}, 08:\penalty0 003, 2005.
\newblock \doi{10.1088/1475-7516/2005/08/003}.

\bibitem[Weinberg(2008)]{Weinberg2008}
Steven Weinberg.
\newblock \emph{Cosmology}.
\newblock Oxford University Press, Oxford, 2008.

\bibitem[Bringmann et~al.(2026)Bringmann, Croon, and
  Sevillano~Mu{\~n}oz]{Bringmann2025}
Torsten Bringmann, Djuna Croon, and Sergio Sevillano~Mu{\~n}oz.
\newblock Updated constraints on the primordial power spectrum at sub-mpc
  scales.
\newblock \emph{Phys. Rev. Lett.}, 137:\penalty0 061002, 2026.

\bibitem[Carr et~al.(2017)Carr, Raidal, Tenkanen, Vaskonen, and
  Veerm{\"a}e]{Carr2017}
Bernard Carr, Martti Raidal, Tommi Tenkanen, Ville Vaskonen, and Hardi
  Veerm{\"a}e.
\newblock Primordial black hole constraints for extended mass functions.
\newblock \emph{Phys. Rev. D}, 96:\penalty0 023514, 2017.

\bibitem[Fixsen et~al.(1996)Fixsen, Cheng, Gales, Mather, Shafer, and
  Wright]{Fixsen1996}
D.~J. Fixsen, E.~S. Cheng, J.~M. Gales, J.~C. Mather, R.~A. Shafer, and E.~L.
  Wright.
\newblock The cosmic microwave background spectrum from the full cobe firas
  data set.
\newblock \emph{Astrophys. J.}, 473:\penalty0 576, 1996.

\bibitem[Hu and Silk(1993)]{HuSilk1993}
Wayne Hu and Joseph Silk.
\newblock Thermalization and spectral distortions of the cosmic background
  radiation.
\newblock \emph{Phys. Rev. D}, 48:\penalty0 485, 1993.

\bibitem[Hu et~al.(1994)Hu, Scott, and Silk]{HuScottSilk1994}
Wayne Hu, Douglas Scott, and Joseph Silk.
\newblock Power spectrum constraints from spectral distortions in the cosmic
  microwave background.
\newblock \emph{Astrophys. J. Lett.}, 430:\penalty0 L5, 1994.

\bibitem[Chluba(2016)]{Chluba2016}
Jens Chluba.
\newblock Which spectral distortions does {$\Lambda$CDM} actually predict?
\newblock \emph{Mon. Not. R. Astron. Soc.}, 460:\penalty0 227, 2016.
\newblock \doi{10.1093/mnras/stw945}.

\bibitem[Bird et~al.(2011)Bird, Peiris, Viel, and Verde]{Bird2011}
Simeon Bird, Hiranya~V. Peiris, Matteo Viel, and Licia Verde.
\newblock Minimally parametric power spectrum reconstruction from the lyman
  $\alpha$ forest.
\newblock \emph{Mon. Not. R. Astron. Soc.}, 413:\penalty0 1717, 2011.

\bibitem[Inomata and Nakama(2019)]{InomataNakama2019}
Keisuke Inomata and Tomohiro Nakama.
\newblock Gravitational waves induced by scalar perturbations as probes of the
  small-scale primordial spectrum.
\newblock \emph{Phys. Rev. D}, 99:\penalty0 043511, 2019.

\bibitem[Agazie et~al.(2023)]{NANOGrav2023}
Gabriella Agazie et~al.
\newblock The nanograv 15 yr data set: Evidence for a gravitational-wave
  background.
\newblock \emph{Astrophys. J. Lett.}, 951:\penalty0 L8, 2023.

\bibitem[Harris et~al.(2020)]{NumPy}
Charles~R. Harris et~al.
\newblock Array programming with numpy.
\newblock \emph{Nature}, 585:\penalty0 357, 2020.

\bibitem[Virtanen et~al.(2020)]{SciPy}
Pauli Virtanen et~al.
\newblock Scipy 1.0: fundamental algorithms for scientific computing in python.
\newblock \emph{Nature Methods}, 17:\penalty0 261, 2020.

\end{thebibliography}

\end{document}